\documentclass[prl,amsmath,twocolumn,showpacs]{revtex4}
\usepackage{times}
\usepackage{graphicx}
\usepackage{amsmath}
\usepackage{mathptmx}

\newcommand{\parti}[2]{\frac{\partial #1}{\partial #2}}

\newcommand{\bra}[1]{\langle#1|}
\newcommand{\ket}[1]{|#1\rangle}

\newcommand{\avg}[1]{\left\langle #1 \right\rangle}

\begin{document}

\title{Beating the Spatial Standard Quantum Limits via Adiabatic
Soliton Expansion}
\author{Mankei Tsang}
\email{mankei@optics.caltech.edu}
\date{\today}
\affiliation{
Department of Electrical Engineering, 
California Institute of Technology, Pasadena, California 91125}
\begin{abstract}
Spatial quantum enhancement effects are studied under a unified
framework.  It is shown that the multiphoton absorption rate of
photons with a quantum-enhanced lithographic resolution is reduced,
not enhanced, contrary to popular belief. The use of adiabatic soliton
expansion is proposed to beat the standard quantum limit on the
optical beam displacement accuracy, as well as to engineer an
arbitrary multiphoton interference pattern for quantum
lithography. The proposed scheme provides a conceptually simple method
that works for an arbitrary number of photons.
\end{abstract}
\pacs{42.50.Dv, 42.50.St, 42.65.Tg}

\maketitle

In many optical imaging applications, such as atomic force microscopy
\cite{putman} and nanoparticle detection \cite{kamimura}, precise
measurements of the displacement of an optical beam are required. It
is hence important to know what the fundamental limit on the accuracy
of such measurements is placed by the laws of physics, and how one can
approach this limit in an experiment. It is now known that if an
optical beam consists of $N$ independent photons with wavelength
$\lambda$, then the minimum uncertainty in its spatial displacement is
on the order of $\lambda/\sqrt{N}$, the so-called standard quantum
limit \cite{barnett}. The ultimate uncertainty permissable by quantum
mechanics, however, is smaller than the standard quantum limit by
another factor of $\sqrt{N}$ \cite{barnett}. An experiment that beats
this standard quantum limit with nonclassical multimode light has
already been demonstrated \cite{treps}. On the other hand, in other
optical imaging applications, such as lithography, microscopy, and
data storage, detection of extremely small features of an object is
desired. The feature size of an optical intensity pattern cannot be
smaller than $\lambda$, due to the resolution limit
\cite{bornwolf}. Multiphoton absorption allows detection of smaller
feature sizes, and the minimum feature size of multiphoton absorption
using a classical coherent light source is on the order of
$\lambda/\sqrt{N}$ \cite{boto}, which can be regarded as the standard
quantum limit on the multiphoton absorption feature size. Nonclassical
light sources allow one to do better, and the ultimate limit is
smaller than the standard one by another factor of $\sqrt{N}$
\cite{boto,bjork}.  A proof-of-concept experiment of this resolution
enhancement has also been demonstrated \cite{dangelo}.  In the time
domain, very similar quantum limits on the position accuracy of an
optical pulse can be derived \cite{giovannetti_nature}.  Given the
striking similarities among the spatiotemporal quantum limits, one
expects them to be closely related to each other, yet the formalisms
used to described each of them are vastly different
\cite{barnett,boto,bjork,giovannetti_nature}, so a more general
formalism applicable to all spatiotemporal domains would greatly
facilitate our understanding towards the spatiotemporal quantum
enhancement effects.

In this Letter, we apply the temporal formalism
used by Giovannetti \textit{et al.}\ \cite{giovannetti_nature} to the
spatial domain, and show that the uncertainty in the beam displacement
and the spot size of multiphoton absorption are in fact closely
related. Using this newly derived result, we demonstrate how arbitrary
multiphoton interference patterns can arise from a continuous
superposition of coincident-momentum states. We further present an
unfortunate result, namely, that the multiphoton absorption rate is
reduced if the quantum lithography resolution is enhanced, contrary to
popular belief \cite{boto}. Finally, we take advantage of the general
spatiotemporal framework to show that the idea of adiabatic soliton
expansion, previously proposed to beat the temporal standard quantum
limit \cite{tsang_prl}, can also be used to beat the standard
quantum limit on the beam displacement accuracy, as well as generate
an arbitrary multiphoton interference pattern, for an arbitrary
number of photons. The use of solitons is an attractive alternative to
the more conventional use of second-order nonlinearity for quantum
information processing, because the soliton effect bounds the photons
together and allows a much longer interaction length for significant
quantum correlations to develop among the photons.

Consider $N$ photons with the same frequency $\omega$ and
polarization that propagate in the $x-z$ plane.
A general wavefunction that describes such
photons is given by \cite{mandel}
\begin{align}
\ket{\Psi} &= \frac{1}{\sqrt{N!}}\int dk_1dk_2...dk_N
\mbox{ }
\phi(k_1,k_2,...,k_N)\ket{k_1,k_2,...,k_N}.
\end{align}
where $\ket{k_1,...,k_N}$ is the momentum eigenstate, $k_1,...,k_N$
specify the transverse wave vectors of the photons along the $x$ axis,
and $\phi(k_1,...,k_N)$ is defined as the multiphoton momentum probability
amplitude. The longitudinal wave vectors along the $z$ axis are all assumed to
be positive. Due to the resolution limit, $\phi(k_1,...,k_N)$ is
band-limited, i.e., $\phi(k_1,...,k_N) = 0$ for any $|k_i| > 2\pi/\lambda$.
One can then define the corresponding quantities in real space,
\begin{align}
&\quad \ket{x_1,...,x_N} 
\nonumber\\
&= \int \frac{dk_1}{\sqrt{2\pi}}...\frac{dk_N}{\sqrt{2\pi}}
\exp(-ik_1x_1-...-ik_Nx_N)\ket{k_1,...,k_N},
\\
&\quad\psi(x_1,...,x_N)
\nonumber\\
&= \int \frac{dk_1}{\sqrt{2\pi}}...\frac{dk_N}{\sqrt{2\pi}}
\phi(k_1,...,k_N)\exp(ik_1x_1+...+ik_Nx_N),
\\
&\quad \ket{\Psi}= \frac{1}{\sqrt{N!}}\int dx_1...dx_N
\psi(x_1,...,x_N)\ket{x_1,...,x_N},
\end{align}
where $\psi(x_1,...,x_N)$ is the multiphoton spatial probability amplitude.
$\phi$ and $\psi$ are subject to normalization conditions $\int
dk_1...dk_N |\phi|^2 = \int dx_1...dx_N |\psi|^2 = 1$, and $\phi$ and
$\psi$ must be symmetric under any exchange of labels due to the bosonic
nature of photons.  The magnitude squared of $\psi$ gives the joint
probability distribution of the positions of the photons,
\begin{align}
\avg{:I(x_1)...I(x_N):} &\propto
\frac{1}{N!}\bra{\Psi}\hat{A}^\dagger(x_1)...\hat{A}^\dagger(x_N)
\hat{A}(x_1)...\hat{A}(x_N)\ket{\Psi}
\\
&= |\psi(x_1,...,x_N)|^2,
\end{align}
where $\hat{A}(x_i)$ and $\hat{A}^\dagger(x_i)$ are the spatial
annihilation and creation operators respectively.  The statistical
interpretation of $\psi$ is valid because we only consider photons
that propagate in the positive $z$ direction.  The above definition of
a multiphoton state is more general than those used by other authors,
in the sense that we allow photons with continuous momenta, compared
with the use of only one even spatial mode and one odd mode by Fabre
\textit{et al.}\ \cite{barnett}, the use of only two discrete momentum
states by Boto \textit{et al.}\ \cite{boto}, and the use of many
discrete momentum states by Bj\"ork \textit{et al.}\ \cite{bjork}.

The displacement of an optical beam can be represented by the
operator
$\hat{X} = (1/N)\int dx\mbox{ }x \hat{A}^\dagger(x)\hat{A}(x)$.
Applying $\hat{X}$ to $\ket{x_1,...,x_N}$ gives
\begin{align}
\hat{X}\ket{x_1,...,x_N} &=\left(\frac{1}{N}\sum_{i=1}^N x_i\right)
\ket{x_1,...,x_N},
\end{align}
so the beam displacement is, intuitively, the mean position
of the photons under the statistical interpretation. If we
assume that $\avg{\hat{X}} = 0$ for simplicity, the 
root-mean-square displacement uncertainty is given by
\begin{align}
\Delta X &\equiv \avg{\hat{X}^2}^{1/2}
\\
&=\left[\int dx_1...dx_N
\left(\frac{1}{N}\sum_{i=1}^N x_i\right)^2
|\psi(x_1,...,x_N)|^2\right]^{1/2}.
\end{align}
It is often more convenient to use a different system of coordinates
as follows \cite{hagelstein},
\begin{align}
X = \frac{1}{N}\sum_{i=1}^N x_i,
\quad
\xi_i = x_i-X,
\mbox{ }
i = 1,...,N-1,
\quad
\xi_N = -\sum_{i=1}^{N-1}\xi_i.
\label{centerofmass}
\end{align}
$X$ is therefore the ``center-of-mass'' coordinate that characterizes
the overall displacement of the optical beam, and $\xi_i$'s are
relative coordinates.
Defining a new probability amplitude in terms of these coordinates,
\begin{align}
\psi'(X,\xi_1,...,\xi_{N-1}) &=
\psi(X+\xi_1,...,X+\xi_N),
\end{align}
we obtain the following expression for the displacement uncertainty,
\begin{align}
\Delta X &= \left[\frac{1}{N}\int dXd\xi_1...\xi_{N-1}
\mbox{ }X^2|\psi'(X,\xi_1,...,\xi_{N-1})|^2\right]^{1/2},
\end{align}
which is the \emph{marginal} width of $\psi'$ with respect to $X$.

On the other hand, the dosing operator of $N$-photon absorption is
given by
\begin{align}
\avg{:I^N(x):} \propto |\psi(x,x,...,x)|^2
=|\psi'(x,0,...,0)|^2,
\end{align}
which is, intuitively, the probability distribution of all $N$ photons
arriving at the same place $x$.
Hence, designing a specified multiphoton interference pattern in
quantum lithography is equivalent to engineering the \emph{conditional}
probability distribution $|\psi'(X,0,...,0)|^2$.

In particular, the
spot size of multiphoton absorption is the conditional width of
$\psi'$ with respect to $X$,
\begin{align}
\left[\int dx\mbox{ }x^2\avg{:I^N(x):}\right]^{1/2}
&\propto \left[\int dX\mbox{ } X^2 |\psi'(X,0,...,0)|^2 \right]^{1/2}
\\
&=
\Delta X \big|_{\xi_1=...=\xi_{N-1}=0}.
\end{align}
Despite the subtle difference between the marginal width
and the conditional width, if $\psi'$ can be made separable in the
following way,
\begin{align}
\psi'(X,\xi_1,...,\xi_{N-1}) &= \bar\psi(X)\psi_{rel}(\xi_1,...,\xi_{N-1}),
\label{separable}
\end{align}
then both widths are identical, and one can optimize
the multiphoton state simultaneously for both applications.

The standard quantum limit on the uncertainty in $X$ is obtained when
the photons are spatially independent, such that $\psi(x_1,...,x_N) =
f(x_1)...f(x_N)$. For example, if $f(x)$ is a Gaussian given by $f(x)
\propto \exp\left(-\kappa^2 x^2/2\right)$, then both the
marginal and conditional uncertainties in $X$ are
\begin{align}
\Delta X_{SQL} = \Delta X_{SQL}
\big|_{\xi_1=...=\xi_{N-1}=0}=
\frac{1}{\sqrt{2N}\kappa}.
\label{sql}
\end{align}
Similar to the optimization of temporal position accuracy
\cite{giovannetti_nature}, the ultimate quantum limits on spatial
displacement accuracy and quantum lithography feature size
are achieved with the following nonclassical state,
\begin{align}
\ket{\Psi} &= \int dk G(k)\ket{k,k,...,k}.
\label{cat}
\end{align}
The momentum probability amplitude is then
\begin{align}
\phi(k_1,...,k_N) &= G(k_1)\delta(k_1-k_2)\delta(k_1-k_3)...\delta(k_1-k_N),
\label{coincidentp}
\end{align}
which characterizes $N$ photons with coincident momentum.
The spatial amplitude is thus given by
\begin{align}
\psi'(X,\xi_1,...,\xi_{N-1}) &= \int \frac{dk}{\sqrt{2\pi}} G(k)\exp(iNkX)
\equiv g(NX),
\label{spatialcat}
\end{align}
which is a function of $X$ only and can be understood as a continuous
superposition of $N$-photon coincident-momentum eigenstates, each with
an effective de Broglie wavelength equal to $2\pi/(Nk)$.  This representation
is equivalent to Boto \textit{et.\ al.}'s proposal \cite{boto}, when
$G(k) \sim \delta(k-k_0) + \delta(k+k_0)$, where $k_0$ is the transverse
wave vector of either arm of the interferometric scheme. The
multiphoton interference pattern is therefore trivially given by
$|g(NX)|^2$, the magnitude squared of the Fourier transform of
$G(k)$. An arbitrary interference pattern can hence be generated, if
an appropriate $G(k)$ can be engineered.  This approach of designing
the multiphoton interference pattern should be compared with the less
direct approaches by the use of discrete momentum states
\cite{boto,bjork}.  With the resolution limit, $G(k) = 0$ for
$|k|>2\pi/\lambda$, so, given the Fourier transform relation between
$G(k)$ and $g(NX)$, the minimum feature size of multiphoton
interference is on the order of $\lambda/N$.

To compare the ultimate uncertainty in $X$ with the standard
quantum limit, let $G(k)$ be a Gaussian given by
$G(k) \propto \exp\left[-k^2/(2\kappa^2)\right]$,
then the uncertainty in $X$ becomes
\begin{align}
\Delta X_{UQL} &= \frac{1}{\sqrt{2}N\kappa},
\end{align}
which is smaller than the standard quantum limit, Eq.~(\ref{sql}), by
another factor of $\sqrt{N}$, as expected.

Let us recall Boto \textit{et al.}'s heuristic argument concerning the
multiphoton absorption rate of entangled photons. They claim that,
because entangled photons tend to arrive at the same place at the same
time, the multiphoton absorption rate must be enhanced \cite{boto}.
If photons tend to arrive at the same place, then the uncertainties
$\Delta \xi_i$ in their relative positions $\xi_i$ must be
small. However, the spatial probability amplitude that achieves the
ultimate lithographic resolution, Eq.~(\ref{spatialcat}), is a
function of $X$ only, which means that the uncertainties in $\xi_i$
are actually infinite. In general, any enhancement of resolution with
respect to $X$ must result in increased uncertainties in the relative
positions $\xi_i$, in order to maintain the same maximum bandwidth. Hence,
Boto \textit{et al.}'s argument, as far as the spatial domain is
concerned, manifestly does not hold for photons with a
quantum-enhanced lithographic resolution. In fact, the opposite is
true: although these photons have a reduced uncertainty in their
average position $X$, they do not arrive at the same place as
often, and the multiphoton absorption rate must be reduced.

To observe this fact, consider the total multiphoton absorption rate
\begin{align}
\int dx \avg{:I^N(x):} \propto \int dX |\psi'(X,0,...,0)|^2.
\end{align}
Because $\psi'$ must satisfy the normalization condition,
\begin{align}
\frac{1}{N}\int dXd\xi_1...d\xi_{N-1} |\psi'(X,\xi_1,...,\xi_{N-1})|^2=1,
\end{align}
$\int
dX |\psi'(X,0,...,0)|^2$ is inversely proportional to $(\Delta
\xi_i)^{N-1}$. An increase in each $\Delta \xi_i$ by a factor of
$\gamma$ hence reduces the total absorption rate by a factor of
$\gamma^{N-1}$.

With all that said, if the multiphoton absorption rate
is reduced due to a quantum-enhanced lithographic resolution, one can
still compensate for this rate reduction by reducing the relative
\emph{temporal} position uncertainties of the photons
\cite{javanainen}.

We now turn to the problem of producing the nonclassical multiphoton
states for spatial quantum enhancement by the use of optical solitons.
Consider the Hamiltonian that describes the one-dimensional
diffraction effect and Kerr nonlinearity on an optical beam
in a planar waveguide,
\begin{align}
\hat{H} &= \int dx \left[-b\parti{\hat{A}^\dagger}{x}\parti{\hat{A}}{x}
+c\hat{A}^\dagger\hat{A}^\dagger\hat{A}\hat{A}
\right],
\end{align}
where $b$ is the Fresnel diffraction coefficient, assumed to be
positive, and $c$ is the negative Kerr coefficient, assumed to be
negative, so that $b/c < 0$ and solitons can exist under the
self-focusing effect. The soliton solution of the spatial amplitude
for $N$ photons under this Hamiltonian is \cite{lai}
\begin{align}
\psi&=
C\int \frac{dk}{\sqrt{2\pi}}
G(k)\exp\bigg[ik\sum_i x_i+\frac{c}{2b}\sum_{i<j}|x_i-x_j|
\nonumber\\&\quad
-ibNk^2t+i\frac{c^2}{12b}N(N^2-1)t\bigg],
\end{align}
where $C = \sqrt{(N-1)!|c/b|^{N-1}/(2\pi)}$ and
$G(k)$ is determined by the initial conditions. If initially the
photons are uncorrelated, $G(k)$ can be approximated as
\cite{tsang_prl}
\begin{align}
G(k) \propto \exp\Big(-\frac{k^2}{2\kappa^2}\Big),
\quad
\kappa = \bigg|\sqrt{\frac{N}{4q}}\frac{c}{b}\bigg| \sim
\frac{1}{\sqrt{N}W_0},
\end{align}
where $q$ is a parameter on the order of unity \cite{tsang_prl}, and
$W_0$ is the initial soliton beam width.  The probability amplitude can
be written in terms of the center-of-mass and relative coordinate
system defined in Eqs.~(\ref{centerofmass}) as
\begin{align}
\psi'&=C\int \frac{dk}{\sqrt{2\pi}}
G(k)\exp\bigg[iNkX+\frac{c}{2b}\sum_{i<j}|\xi_i-\xi_j|
\nonumber\\&\quad
-ibNk^2t+i\frac{c}{12b}N(N^2-1)t\bigg],
\end{align}
which is separable in the way described by Eq.~(\ref{separable}),
meaning that the conditional width and marginal width with respect to
$X$ are identical.

If we adiabatically reduce $c$ or increase $b$
along the waveguide by, for example, increasing the waveguide modal
thickness, then we can reduce the uncertainty in the relative momenta of the
photons and increase the uncertainty in the relative positions
\cite{tsang_prl}. Classically, we expect the soliton beam width to
expand and the spatial bandwidth to be reduced, But the most crucial
difference in the quantum picture is that the center-of-mass
coordinate $X$ remains unaffected during the adiabatic soliton
expansion, apart from the quantum dispersion term $-ibNk^2t$.
In the limit of vanishing $c/b$, the wavefunction
approaches the ultimate multiphoton state given by
Eqs.~(\ref{cat}), (\ref{coincidentp}), and (\ref{spatialcat}).

As pointed out in Ref.~\cite{tsang_prl}, the quantum dispersion term
can be compensated if the soliton propagates in a second medium with
an opposite diffraction coefficient $b'$. Full compensation is
realized when $\int_0^T b(t) dt = -\int_0^{T'} b'(t) dt$, where $T$ is
the propagation time in the first medium and $T'$ is the propagation
time in the second medium.  Negative refraction \cite{pendry}
is hence required in the second medium.  Ideally the
second medium should also have a Kerr coefficient $c'$ opposite to the
final value of $c$ in the first medium, such that $c' = -c(T)$, so
that $b'/c'<0$ and the quantum soliton maintains its shape, but in
practice $c'=0$ also suffices, because the multiphoton spectrum
$|\phi|^2$ remains unchanged in a linear dispersive medium while
the quantum dispersion is being compensated.

\begin{figure}[htbp]
\centerline{\includegraphics[width=0.48\textwidth]{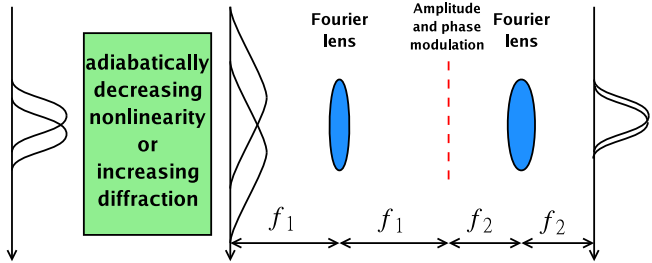}}
\caption{(color online).
Schematics of the spatial quantum enhancement setup via
adiabatic soliton expansion.}
\label{adiabatic}
\end{figure}

So far we have worked in the paraxial regime, so a 4$f$ imaging system
with spatial phase modulation in the Fourier plane can effectively
mimic the behavior of negative refraction. Consider the system
depicted in Fig.~\ref{adiabatic}.  After the beam goes through the
nonlinear medium and the first Fourier lens, in the Fourier plane, the
multiphoton amplitude becomes
\begin{align}
\psi(x_1,...,x_N) &\propto
\phi\left(\frac{2\pi x_1}{\lambda f_1},...,\frac{2\pi x_N}{\lambda f_1}\right).
\end{align}
A quadratic spatial phase modulation in the Fourier plane, by a
lens for example, can hence act as negative refraction in the paraxial
regime and cancel the quantum dispersion term.

In general, Fourier-domain modulation with a transfer function $H(k)$
can be used to shape $G(k)$, if the ultimate multiphoton state is
achieved, because all photons have coincident momenta, resulting in
an output given by $G(k)H^N(k)$. A desired multiphoton interference
pattern $|g(NX)|^2$ can hence be engineered by spatial modulation in
the Fourier plane.

If the ultimate multiphoton state is achieved via adiabatic soliton
expansion, the bandwidth of the optical beam is the same as that of
$G(k)$, which is on the order of $1/(\sqrt{N}W_0)$, and is much
smaller than the resolution limit. The second Fourier lens in
Fig.~\ref{adiabatic} should therefore have a small focal length $f_2$
to demagnify the optical beam and increase the bandwidth of $G(k)$.

Current technology should be able to expand a spatial
soliton with $N\sim 10^{10}$ photons by a few times before decoherence
effects, such as loss, become significant.
However, for quantum lithography, an ideal high-number-photon
absorption material is difficult to obtain, so a giant Kerr
nonlinearity, such as that theorized in a coherent medium
\cite{schmidt}, is required to produce a few-photon soliton.

In conclusion, spatial quantum enhancement effects are studied under a
general framework. It is shown that the multiphoton absorption rate is
reduced if the lithographic resolution is enhanced. The use of
adiabatic soliton expansion is proposed to
beat the spatial standard quantum limits for an arbitrary number of
photons.

Discussions with Demetri Psaltis and financial
support by the Defense Advanced Research Projects Agency (DARPA)
are gratefully acknowledged.

\end{document}